\documentclass[
floatfix,
reprint,
 amsmath,amssymb,
 aps,
]{revtex4-2}
\usepackage{comment}
\usepackage{amsmath}
\usepackage[nolist]{acronym}
\usepackage{float}

\usepackage{tcolorbox}

\usepackage{graphicx}
\usepackage{dcolumn}
\usepackage{bm}

\begin{document}

\title{Nonlocal current-driven heat flow in ideal plasmas}
\author{Nicholas Mitchell$^1$}
\author{David Chapman$^2$}
\author{Grigory Kagan$^1$}%
\affiliation{$^1$The Blackett Laboratory, Imperial College, London SW7 2AZ, UK
}%
\affiliation{$^2$First Light Fusion Ltd., Unit 9/10 Oxford Pioneer Park, Mead Road, Yarnton, Kidlington OX5 1QU, UK
}%

\date{\today}

\begin{abstract}
    Electron heat flux is an important and often dominant mechanism of energy transport in a variety of collisional plasmas in a confined fusion or astrophysical context. While nonlocal conductive heat transport, driven by strong temperature gradients, has been investigated extensively in previous literature, nonlocal regimes of the current-driven heat flow and friction have not received the same attention. In this work, a first-principles reduced kinetic method (RKM) is applied to study nonlocal effects on current-driven transport. In addition to nonlocality due to sharp gradients, sufficiently large currents are found to significantly enhance current-driven heat flux due to a novel nonlocal mechanism, with this enhancement being increasingly prevalent for higher effective ionizations $Z^*$. Introducing the dimensionless number $N_u \equiv \vert \boldsymbol{u}_e - \boldsymbol{u}_i \vert / v_{\text{th},e}$, these enhancements occur for even relatively weak flows $N_u \gtrsim 1/100$, analogously to standard nonlocal effects becoming significant for Knudsen numbers $N_K \gtrsim 1/100$.
\end{abstract}

\maketitle

\begin{acronym}
\acro{VFP}{Vlasov-Fokker-Planck}
\acro{MFP}{mean free path}
\acro{BGK}{Bhatnagar–Gross–Krook}
\acro{DF}{distribution function}
\acro{DDF}{deviation of distribution function from Maxwellian}
\acro{CoM}{centre-of-momentum}
\acro{RKM}{reduced kinetic method}
\acro{LTE}{local thermal equilibrium}
\acro{ICF}{inertial confinement fusion}
\end{acronym}

It has long been established that in a wide range of practically important scenarios the electron heat conduction becomes nonlocal even when the plasma appears close to \ac{LTE} \cite{PhysRevLett.46.243,JDCallen_1997,Rinderknecht_2018,Power_2023,PhysRevLett.92.205006,Chen_2024,lau2025nonlocaleffectsthermaltransport,10.1063/5.0018733}. Standard Chapman-Enskog type calculations \cite{chapman_cowling,Spitzer1962,Braginskii1965ReviewsOP,10.1063/1.865901,10.1063/1.4801022} yield a local analytic expression for the electron conductive heat flux $\boldsymbol{q}^T_e\propto-\boldsymbol{\nabla}T_e$ that breaks down even for seemingly small Knudsen numbers \cite{Epperlein91}. This Knudsen number is conventionally defined as $N_K = \lambda_{\text{th}}/L$, where $\lambda_{\text{th}}$ is the thermal \ac{MFP} and $L \sim |\boldsymbol{\nabla} T_e/T_e|^{-1}$ is the gradient length scale of the electron temperature. The ultimate reason for the conductive heat flux becoming nonlocal even for seemingly smooth temperature profiles is that heat is transported predominantly by electrons with velocities $ v \lesssim 4 v_{\text{th},e}$, where $v_{\text{th},e}=\sqrt{2 k_{\text{B}} T_e / m_e}$ is the electron thermal velocity. In plasmas, a charged particle's \ac{MFP} scales as the fourth power of the particle velocity $\lambda_{\text{th}} \propto v^4$, so the \ac{MFP}s of electrons relevant to heat transport can be comparable to $L$ even if $N_K \ll 1$, leading to nonlocal deviations of $\boldsymbol{q}_e^T$. 

Owing to large practical importance of heat transport in plasmas, extensive literature focuses on the conductive electron heat flux. The physics behind its nonlocality is well understood, as outlined above; first-principles studies have been carried out \cite{Epperlein91,10.1063/5.0086783,Bell_2024,THOMAS20121051} and computationally efficient nonlocal closures have been developed \cite{SNB_model,10.1063/1.2179392,10.1063/1.4926824,Brodrick_2017,10.1063/5.0250147,e_heat_conduction_icf,10.1063/1.5011818}. What has largely been overlooked, however, is the nonlocal physics of the current-driven heat flow, in the plasma context known as  $\boldsymbol{q}^u_e \propto \Delta\boldsymbol{u} \equiv \boldsymbol{u}_e - \boldsymbol{u}_i = -\boldsymbol{j}/en_e $ of Braginskii's seminal monograph \cite{Braginskii1965ReviewsOP}, which describes the Peltier and Ettingshausen effects. Although considered in a few works, nonlocal modifications to $\boldsymbol{q}^u_e$ have been attributed to the same spatial scales based on the Knudsen number \cite{10.1063/1.2337789,Watkins_thesis}. Since these works relied on simplified rather than first-principles approaches with this assumption essentially built-in, a novel driver of the nonlocal behavior, the electron flow itself, has been missed.

In this Letter, we demonstrate that the dimensionless number $ N_{u} \equiv \vert \Delta \boldsymbol{u}  \vert / v_{\text{th},e} $ plays a role to $\boldsymbol{q}_e^u$ analogous to that $N_K$ plays for $\boldsymbol{q}_e^T$. Although one might expect that the standard local calculation is valid with $N_u\ll1$, a much more restrictive condition is found to be required. Both an intuitive explanation of the underlying transport physics and first-principles kinetic calculations are presented. A similar but smaller effect is also found on the electron-ion friction $\boldsymbol{R}_e^u$. Deviations from standard results due to this effect is increasingly prevalent at higher effective ionizations $Z^*\equiv \langle Z^2\rangle/\langle Z\rangle $ since the relative importance of electron-ion collisions over electron-electron collisions is increased. 

Local heat conduction in a collisional magnetized plasma is widely understood, originating from random motion of electrons that diffuses heat due to a temperature gradient. Analysis of small-angle binary Coulomb collisions shows that the collision frequency of a test electron against an ion background is inversely proportional to the cube of its speed in the frame comoving with the ion population, i.e., $\nu_{ei} \propto 1/ \vert \boldsymbol{v}-\boldsymbol{u}_i\vert^3 $. This fact has the consequence that relative flows, or currents, also drive heat flux as electrons moving in the same direction as the relative drift ($\boldsymbol{w}\cdot\Delta\boldsymbol{u} > 0$, where $\boldsymbol{w} = \boldsymbol{v}-\boldsymbol{u}_e$ is the electron peculiar velocity) are less collisional on average against ions than electrons moving against the relative drift ($\boldsymbol{w}\cdot\Delta\boldsymbol{u} < 0$). The heat carried by these two sides of the electron population then does not cancel and there is a net heat flux in the direction of the relative flow $\boldsymbol{q}_e^u \propto \Delta\boldsymbol{u}$. Currents necessitate the inclusion of magnetic fields which cause electrons to undergo gyro-orbits around the magnetic field lines, suppressing heat transport in the direction perpendicular to the field. However, the gyro-orbits also generate particle motion in the direction perpendicular to both the magnetic field and temperature gradient or relative flow, therefore an additional `cross' component to each of the above transport fluxes is gained in this direction. 

The above picture allows a qualitative explanation as to why local current-driven heat flux results \cite{Spitzer1962,Braginskii1965ReviewsOP,10.1063/1.865901,10.1063/1.4801022} break down even for relatively weak electron flows. A nonvanishing net electron flow implies a nonzero electric field set by the bulk electron distribution. This flow being much slower than $v_{\text{th},e}$ is equivalent to the electric field being much less than the Dreicer field \cite{helander_and_sigmar}. While this ensures the bulk distribution is not in a disruptive runaway scenario, the dynamic friction experienced by suprathermal electrons is lower due to the aforementioned collision frequency scaling. Dynamic friction can then be insufficient to prevent the electric field from driving suprathermal electrons farther away from thermodynamic equilibrium than can be captured by a local transport theory. Since this mechanism accelerates electrons in the direction of the relative flow $\Delta\boldsymbol{u}$, nonlocal behavior is expected to vary for the different magnetized components of current-driven heat flow in the parallel, perpendicular, and cross directions. We now proceed to the quantitative first-principles description of these nonlocal effects.

In a sufficiently collisional simple plasma, the local results \cite{10.1063/1.4801022} for the current-driven electron heat flux and friction force are
\begin{equation}
    \begin{aligned}
         \boldsymbol{q}_e^u &= n_e k_{\text{B}} T_e\big (\underbrace{ (\hat{\beta}_\parallel \boldsymbol{b}\boldsymbol{b} + \hat{\beta}_\perp(\mathbb{I}-\boldsymbol{b}\boldsymbol{b}))\cdot \Delta\boldsymbol{u} }_{\text{Peltier}} + \underbrace{\hat{\beta}_\times \boldsymbol{b}\times \Delta\boldsymbol{u}}_{\text{Ettingshausen}} \big),
        \\ \boldsymbol{R}_e^u &= - \frac{n_e m_e}{\tau_{ee}} \big( \underbrace{ (\hat{\alpha}_\parallel \boldsymbol{b}\boldsymbol{b} + \hat{\alpha}_\perp(\mathbb{I}-\boldsymbol{b}\boldsymbol{b}))\cdot \Delta\boldsymbol{u}}_{ \text{Friction}}- \underbrace{  \hat{\alpha}_\times \boldsymbol{b}\times \Delta \boldsymbol{u}}_{\text{Cross-friction}} \big),
    \end{aligned}
\end{equation}
where $\tau_{ee}$ is the electron-electron collision time and $\boldsymbol{b}=\boldsymbol{B} / \vert \boldsymbol{B}\vert$ is the unit vector parallel to the magnetic field. The dimensionless coefficients $\hat{\alpha}_{\parallel,\perp,\times}$ and $\hat{\beta}_{\parallel,\perp,\times}$ are dependent on the effective ionization $Z^*$ and, for perpendicular and cross terms, the hall parameter $\chi =  \omega_{\text{c},e} \tau_{ee}$ with electron cyclotron frequency $\omega_{\text{c},e} = e\vert \boldsymbol{B} \vert / m_e$.

The \ac{RKM} from previous work \cite{Mitchell_2024,PhysRevLett.115.105002} has been extended to solve the \ac{VFP} equation for the deviation of the electron distribution function from Maxwellian $\delta f_e = f_e - f^M_e$ in 1D3V coordinates ($z$,$\boldsymbol{w}$) given input fluid profiles. The \ac{RKM} evaluates nonlocal transport from the input profiles, in contrast with a full \ac{VFP} simulation which relies on evaluating transport quantities from relaxing profiles, which may have changed significantly from the input profiles before the quasistationary transport fluxes are established. Here, this method is applied to electrons, where the \ac{VFP} equation for $\delta f_e$ in the frame moving with the local electron flow $\boldsymbol{u}_e$ is

\begin{equation}\label{main_electron_equation}
    \begin{aligned}
        &f^M_e \bigg[ \bigg(\boldsymbol{R}_e /p_e  + \bigg( \frac{w^2}{v_{\text{th},e}^2} - \frac{5}{2}\bigg)  \boldsymbol{\nabla} \ln T_e \bigg) \cdot \boldsymbol{w}
        \\& + \frac{2}{v_{\text{th},e}^2} \boldsymbol{\nabla}\boldsymbol{u}_e \colon \bigg(\boldsymbol{w}\boldsymbol{w} - \frac{1}{3}w^2\mathbb{I} \bigg)\bigg] - C_{ei}\{f^M_e \}
        \\&  = C_{ee}\{ \delta f_e \} + C_{ei}\{ \delta f_e \} + \frac{e}{m_e} (\boldsymbol{w} \times \boldsymbol{B}) \cdot\boldsymbol{\nabla}_{\boldsymbol{w}} \delta f_e
        \\& \quad - \bigg[ \boldsymbol{w} \cdot\boldsymbol{\nabla}
        -\bigg(\boldsymbol{w}\cdot\boldsymbol{\nabla}\boldsymbol{u}_e+ \frac{\boldsymbol{R}_e -\boldsymbol{\nabla}p_e}{\rho_e} \bigg)  \cdot\boldsymbol{\nabla}_{\boldsymbol{w}} \bigg]\delta f_e.
    \end{aligned}
\end{equation}
The distribution function $\delta f_e$ is expanded over spherical harmonics in velocity space, leaving a set of PDEs which are solved numerically. The solution for $\delta f_e$ should not perturb the flow which is considered as an input quantity, therefore should satisfy $\int d^3\boldsymbol{w}\ \boldsymbol{w}\delta f_e = 0$, which is enforced as a constraint in the solver. The heat flux is obtained as $\boldsymbol{q}_e = \int d^3\boldsymbol{w}\ \frac{1}{2}m_e w^2 \boldsymbol{w} \delta f_e$. To obtain the solution with a self-consistent momentum exchange $\boldsymbol{R}_e$, the \ac{RKM} solver is run iteratively, passing in the integrated $\boldsymbol{R}_e = \int d^3\boldsymbol{v}\ m_e \boldsymbol{v} C_{ei}\{ f_e \} $ from the previous iteration as the input $\boldsymbol{R}_e$ appearing in Equation \ref{main_electron_equation} until they match. This usually takes only one or two iterations.

To investigate nonlocality of current-driven transport, we consider an isothermal isochoric simple plasma with a flow in the $y$ direction of the form
\begin{equation}
    \Delta \boldsymbol{u} = u_0 \exp( - (z/L)^2)\boldsymbol{\hat{y}},
\end{equation}
with a uniform magnetic field in the $-x$ direction $\boldsymbol{B}=-B_0\boldsymbol{\hat{x}}$ and no ion flow $\boldsymbol{u}_i = 0$. The peak electron flow with $N_u^0 \equiv \vert u_0 \vert/ v_{\text{th},e}  $ describes the maximum speed of the flow relative to the electron thermal speed. The Knudsen number is defined as $N_K \equiv \lambda_{\text{th},e}/L$, where $\lambda_{\text{th},e} = \sqrt{\lambda_{\text{th},ee} \lambda_{\text{th},ei}}  $ is the electron thermal \ac{MFP} \cite{Epperlein91}.
\begin{figure}
    \centering    
    \includegraphics[width=0.49\textwidth]{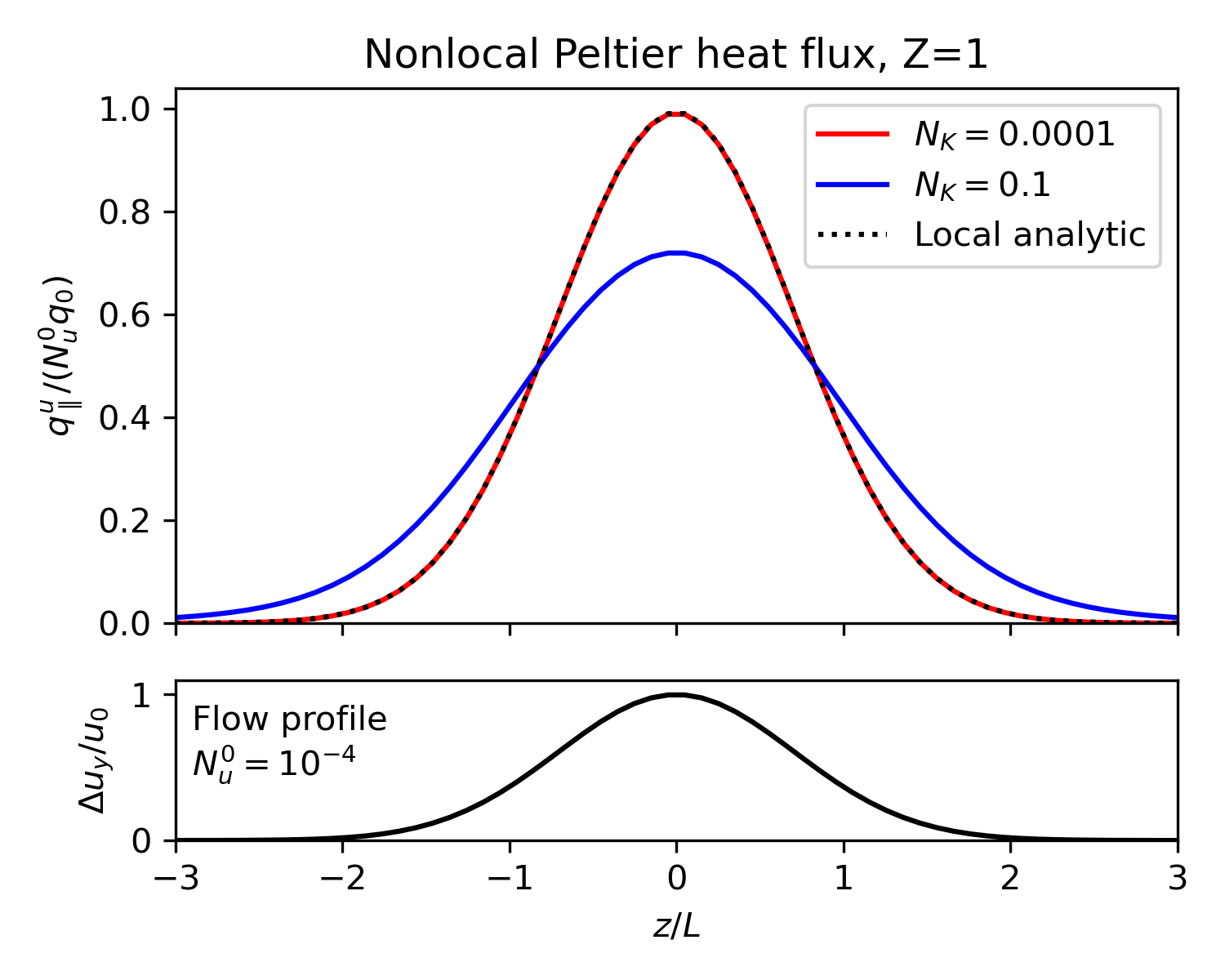}
    \caption{
        Current-driven Peltier heat flux in a simple unmagnetized isothermal isochoric hydrogen ($Z^*=1$) plasma with a peak flow $N_u^0= \vert u_0\vert/v_{\text{th},e} = 10^{-4}$. The heat flux is appropriately normalized against $N_u$ and the free-streaming heat flux $q_0 = n_e k_{\text{B}} T_e v_{\text{th},e}/\sqrt{2}$. For small Knudsen number $N_K=10^{-4}$, the local analytic result for the Peltier heat flux $\boldsymbol{q}_e^u = \hat{\beta}_\parallel n_e k_{\text{B}}T_e \Delta\boldsymbol{u}$, where $\hat{\beta}_\parallel(Z^*=1) \approx 0.71$, is reproduced. For Knudsen number $N_K=0.1$, the Peltier heat flux is noticeably nonlocal, with the peak heat flux reduced and `preheat', where the heat flux away from the peak is enhanced.
    }
    \label{fig:031024_current_low_M}
\end{figure}

The unmagnetized Peltier heat flux $q_{e,\parallel}^u$ in a local and nonlocal case with small peak flow $N_u^0=10^{-4}$ is shown in Figure \ref{fig:031024_current_low_M}. Similarly to the nonlocal conductive heat flux, the peak heat flux is inhibited, and the heat flux is enhanced away from the peak current. For moderate magnetic fields $\chi \sim 1$ and for a wide variety of ionizations $Z^*$, nonlocality of the Peltier heat flux persists. Sufficiently strong magnetic fields $\chi \gg 1$ eventually suppress nonlocality, restoring the heat flux to its local limit even for high Knudsen number. 

Nonlocal deviations conventionally arise from the $\boldsymbol{w}\cdot\boldsymbol{\nabla} \delta f_e$ term in Equation \ref{main_electron_equation}, which affects transport when 
\begin{equation}    \frac{\vert\boldsymbol{w}\cdot \boldsymbol{\nabla} \delta f_e \vert}{\vert C_e\{ \delta f_e\} \vert} \sim \bigg(\frac{w}{v_{\text{th},e}}\bigg)^4 N_K
\end{equation}
is comparable to or greater than unity for particles $w / v_{\text{th},e} \lesssim 4$ that contribute to transport quantities, so where $N_K \gtrsim 1/100$. However, the force term $\boldsymbol{R}_e \cdot \boldsymbol{\nabla}_{\boldsymbol{w}} \delta f_e$ may also cause deviations in $\delta f_e$ and therefore transport from standard results. Ignoring spatial gradients, this term should affect transport where
\begin{equation}
    \frac{\vert \frac{1}{\rho_e} \boldsymbol{R}^u_e\cdot \boldsymbol{\nabla}_{\boldsymbol{w}} \delta f_e \vert}{\vert C_e\{ \delta f_e\} \vert} \sim  \bigg(\frac{w}{v_{\text{th},e}}\bigg)^3 N_u 
\end{equation} 
is comparable to or greater than unity for relevant particle velocities, so where $N_u \gtrsim 1/100$. If all spatial gradient terms are neglected but the force term is included, solutions to $\delta f_e$ cannot satisfy $\delta f_e (w\rightarrow \infty) = 0$ since there always exists a sufficiently large velocity in the tail where electrons become prone to runaway, leading to a divergent solution. However, these runaway electrons in the $w_y$ direction can be depleted by spatial advection in the $z$ direction acting as a sink at large $w$. For sufficiently large length scales in the $z$ direction, this spatial advection can operate at sufficiently high $w$ beyond the velocity space that contributes to the integrand of the heat flux $w \lesssim 4 v_{\text{th},e} $ such that the heat flux is not affected by spatial advection, so remains local. Sufficiently large flows can then cause deviations from traditional local transport calculations due to the force term which rely on spatial advection to be convergent, but yield transport fluxes which are dependent on $z$ only through the local flow $\Delta\boldsymbol{u}$, i.e., $\boldsymbol{q}_e^u = \boldsymbol{q}_e^u(\Delta\boldsymbol{u}(\boldsymbol{x}))$ and similarly for $\boldsymbol{R}_e^u$, therefore are `quasilocal'. Analogously to the Knudsen number $N_K$ characterising the impact of nonlocal effects due to spatial advection from the $\boldsymbol{w}\cdot\boldsymbol{\nabla}$ operator, the number $N_u$ characterises the impact of quasilocal effects due to velocity-space advection from the $\boldsymbol{R}^u_e\cdot\boldsymbol{\nabla}_{\boldsymbol{w}}$ operator. For higher $Z^*$, the friction force against ions increases relative to the energy scattering components of the collision operator, which come only from electron-electron collisions since $m_i \gg m_e$. The quasilocal effect should then be more prevalent in higher $Z^*$ plasmas. 

\begin{figure}
    \centering
    \includegraphics[width=0.48\textwidth]{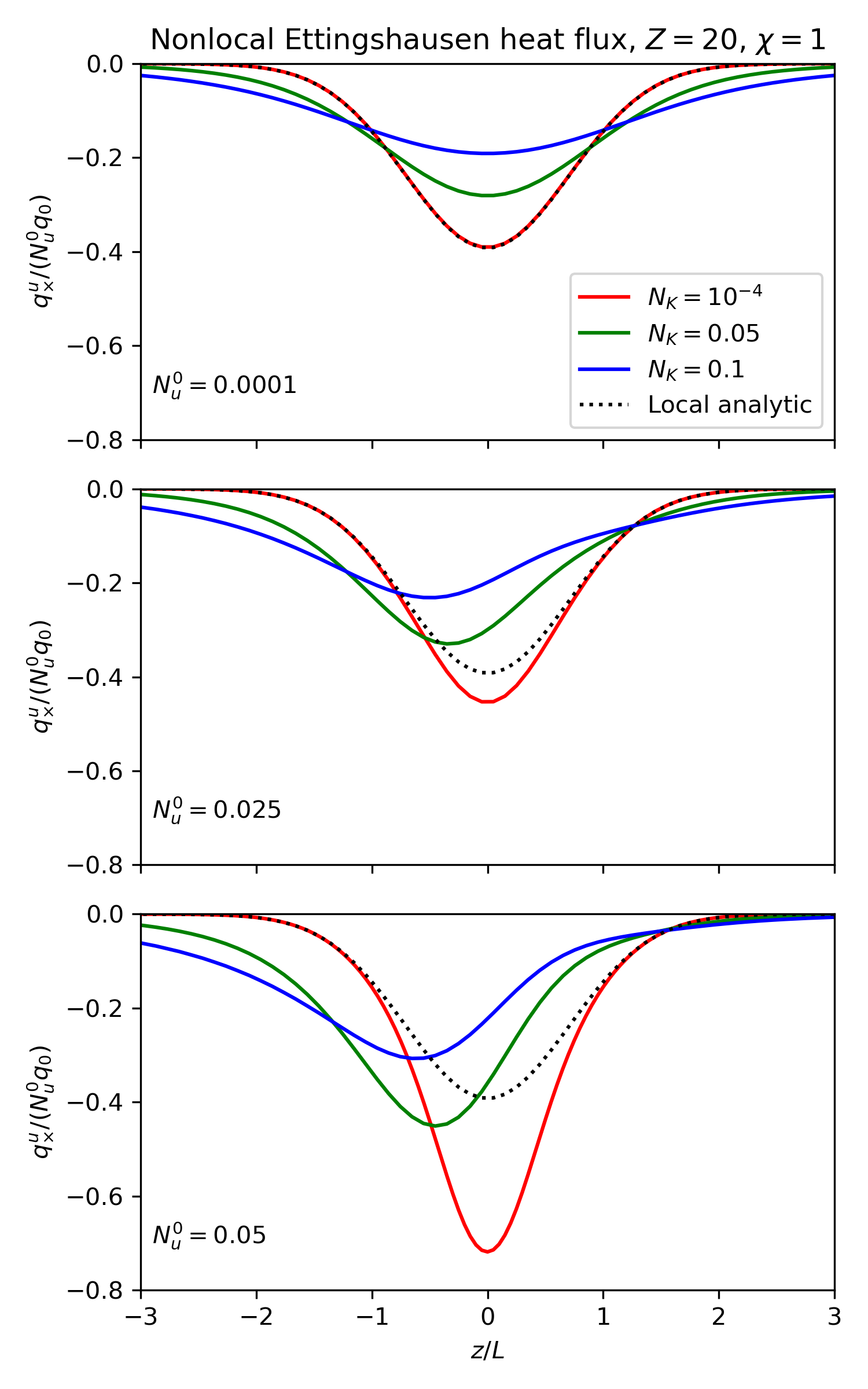}
    \caption{
        Ettingshausen heat flux in a magnetized isothermal isochoric simple plasma with effective ionization $Z^*=20$ with the same shape of flow profile as in Figure \ref{fig:031024_current_low_M}. A uniform magnetic field in the $-x$ direction is chosen $\boldsymbol{B}=-B_0 \boldsymbol{\hat{x}}$ with a field strength that results in a hall parameter $\chi=1$. Different subplots correspond to different peak flows $N_u^0 = 0.0001,\, 0.025,\, 0.05$ which have a noticeable impact on the Ettingshausen heat flux.
    }
    \label{fig:031024_current_high_M}
\end{figure}

Figure \ref{fig:031024_current_high_M} demonstrates the impact of fast flows on the Ettingshausen heat flux for $Z^*=20$. Typical qualitative nonlocal behaviour is exhibited for low $N_u^0$r. For larger $N_u^0$, the heat flux is noticably modified due to the quasilocal effect arising from the $\boldsymbol{R}_e^u \cdot \boldsymbol{\nabla}_{\boldsymbol{w}} \delta f_e$ term in the kinetic equation. For small Knudsen number $N_K = 10^{-4}$, the heat flux is enhanced at large $N_u$, but remains spatially local. For larger Knudsen number, conventional nonlocal effects make the heat flux profile nonlocal in space, reducing and shifting the peak heat flux to $z<0$, and developing significant preheats in regions of small $N_u$. The friction and cross-friction are not shown here since they have very small $\lesssim 5\%$ deviations from local results. 

\begin{figure}
    \centering   
    \includegraphics[width=0.49\textwidth]{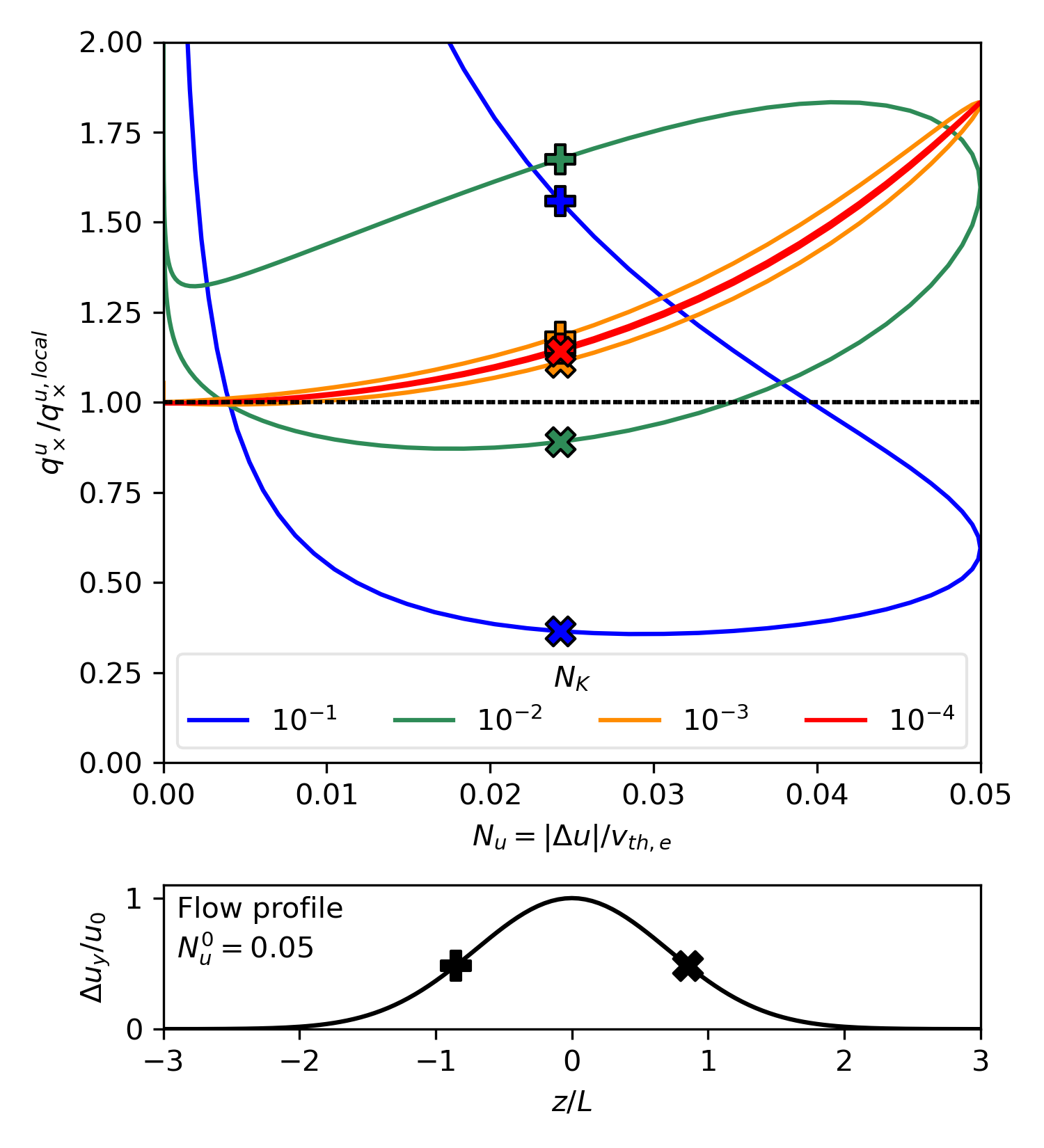}
    \caption{
        Ratio of Ettingshausen heat flux to the local analytic result for peak flow $N_u^0=0.05$, hall parameter $\chi=1$, and ionization $Z^*=20$, plotted as a function of $N_u$ rather than $z/L$ to emphasise the quasilocal effect. A wide variety of Knudsen numbers are chosen to illustrate the transition between the nonlocal and quasilocal regimes. Nonlocal heat fluxes appear as loops on these plots since they are not functions of the local flow. For small Knudsen number, the heat flux tend towards a one-to-one line, showing that the heat flux is local. Plus and cross markers are added to indicate how different spatial positions with the same flow $N_u$ yield the same heat flow for small Knudsen number, but different heat flow for larger Knudsen number. Larger hall parameter $\chi$ suppresses both nonlocal and quasilocal deviations, driving these curves closer to the black dashed line.
    }
    \label{fig:051124_current_high_Z10}
\end{figure}

Figure \ref{fig:051124_current_high_Z10} emphasizes the quasilocal effect by plotting the ratio against the local result as a function of $N_u = \vert \Delta\boldsymbol{u} / v_{\text{th},e} \vert$. The red line for small Knudsen number $N_K=10^{-4}$ is one-to-one and returns to a ratio of unity as $N_u\rightarrow 0$. This indicates that the Ettingshausen heat flux is still local, i.e., $q^u_{e,\times}=q^u_{e,\times}(u(z))$, but deviates away from the standard analytic result at large $N_u\sim 0.01-0.05$ due to the quasilocal effect. The blue and green lines for larger Knudsen numbers form loops, indicating that the heat flux is nonlocal, i.e., $q^u_{e,\times}$ may not be expressed as a function of the local $u$. The plus and cross markers indicate two spatial positions with the same flow. For the nonlocal case, these positions have very different heat fluxes. For the $N_K=10^{-4}$ case, the two markers overlap on the red curve, indicating that these two spatial positions still have the same heat flux. The red curve is then not specific to this exact profile, but universal for profiles with sufficiently small Knudsen number for the given ionization and hall parameter. Different flow profiles with $N_u^0 = 0.05$ have been verified to reproduce the same red curve for sufficiently small Knudsen number.

\begin{figure}
    \centering    
    \includegraphics[width=0.49\textwidth]{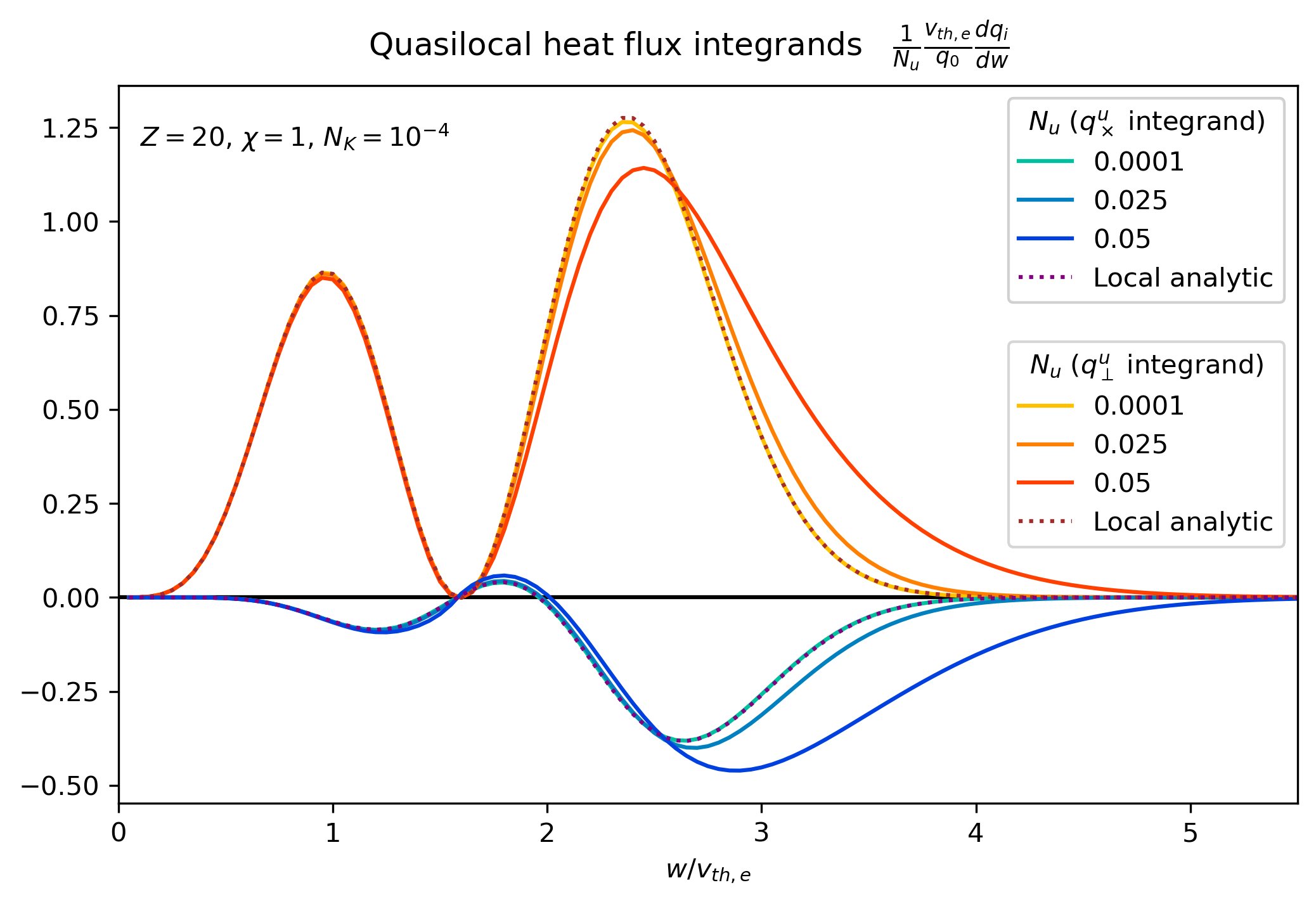}
    \caption{
        Peltier and Ettingshausen heat flux integrands for various flows $N_u$ with ionization $Z^*=20$, hall parameter $\chi=1$, and very small Knudsen number $N_K \lll 1$. For small $N_u=10^{-4}$, the heat flux integrands match up extremely well with the local analytic result. For larger $N_u \sim 0.025 - 0.05$, higher energy electrons $w/v_{\text{th},e} \gtrsim 2.5$ are accelerated, shifting the integrand to larger velocity and enhancing the heat fluxes corresponding to the integrated area beneath the curves. The Ettingshausen integrands are more weighted to higher $w$ than the Peltier integrands, i.e., the Ettingshausen heat flow is carried preferentially by faster particles than the Peltier heat flow, explaining why the Ettingshausen heat flow is more impacted by large $N_u$.
    }
    \label{fig:051124_current_high_Z10_df}
\end{figure}

Figure \ref{fig:051124_current_high_Z10_df} demonstrates the impact of the quasilocal effect on the integrands of the heat flux in the $y$ and $z$ directions. For high $N_u$, the integrand at low velocities $w/v_{\text{th},e} \lesssim 2$ experiences only small deviations due to the $\boldsymbol{R}_e^u \cdot \boldsymbol{\nabla}_{\boldsymbol{w}} \delta f_e$ term since the bulk is highly collisional. In the less-collisional region $2 \lesssim w/v_{\text{th},e} \lesssim 5$, the heat flux integrand for high $N_u$ is greatly enhanced from the local result due to electrons starting to become runaway. Electrons that do become runaway are then accelerated to larger velocities $w/v_{\text{th},e} \gtrsim 5$ where they may be depleted to other regions in physical space by spatial advection, even for very small Knudsen number. This spatial advection occurs in regions of velocity space that do not contribute significantly to the heat flux integrand, thus leaving the heat flux local. Even larger $N_u$, larger $Z^*$, or smaller $\chi$ may then result in divergent integrands, which is indicative of being far from \ac{LTE} where fluid models with quasistatic transport closures are invalid and the plasma requires a full kinetic description.

In summary, current-driven electron heat flux $\boldsymbol{q}_e^u$ becomes nonlocal in the presense of relatively weak spatial gradients with the Knudsen number $N_K \equiv \lambda_{\text{th},e}/L \gtrsim 1/100$, as opposed to $N_K \gtrsim 1$, similarly to the conductive heat flux $\boldsymbol{q}_e^T$. Furthermore, a novel nonlocal modification to $\boldsymbol{q}_e^u$ has been found, which is driven by a relatively weak electron flow $N_u\equiv\vert \boldsymbol{u}_e-\boldsymbol{u}_i\vert/v_{\text{th},e} \gtrsim 1/100$, as opposed to $N_u \gtrsim 1$. The underlying physics has been both captured with a first-principles kinetic model and explained qualitatively and can significantly modify and enhance Ettingshausen heat flow. Such effects may be especially relevant for a variety of collisional plasmas such as plasma staircases \cite{Lopez_Bott_Schekochihin_2025}, dense Z-pinches \cite{JPChittenden_1993,10.1063/5.0082435}, and ICF implosions \cite{walsh_thesis,PhysRevLett.118.155001} where current-driven transport has a notable effect on hydrodynamic evolution and therefore yield. 

\newpage

\bibliography{apssamp}

\end{document}